\documentclass{article}[12pt]
\usepackage{amsfonts}

\newbox\strutbox
\setbox\strutbox=\hbox{\char"B}
\def\strut{\relax\ifmmode\copy\strutbox\else\unhcopy\strutbox\fi}
\def\ialign{\everycr{}\tabskip0pt\halign}
\def\eqalign#1{\null \,\vcenter {\openup\jot \mathsurround 0pt
    \ialign{\strut \hfil$\displaystyle{##}$&$\displaystyle
      {{}##}$\hfil\crcr#1\crcr}}\,}

\def\eqalignno#1{\tabskip 0pt plus 1 fill \halign to\displaywidth{\hfil$\tabskip0pt\everycr{}\displaystyle{##}$\tabskip 0pt &$\tabskip0pt\everycr{}\displaystyle {{}##}$\hfil \tabskip 0pt plus 1 fill&\llap {$\tabskip0pt\everycr{} ##$}\tabskip 0pt \crcr #1\crcr }}

\def \ss{\smallskip}

\begin{document}
\title{Spin-1 Amplitudes in Black-Hole Evaporation}
\author{A.N.St.J.Farley and P.D.D'Eath\protect\footnote{Department of Applied Mathematics and Theoretical Physics,
Centre for Mathematical Sciences,
University of Cambridge, Wilberforce Road, Cambridge CB3 0WA,
United Kingdom}}

\maketitle

\begin{abstract}
This paper extends to Maxwell theory our earlier work, in which we
studied the quantum amplitude (not just the probability) to have final
data $(h_{ij}{\,},\phi)_F$ for Einstein gravity and a massless scalar
field, posed on a final space-like hypersurface $\Sigma_F{\,}$, given
initial data $(h_{ij}{\,},\phi)_I$ on an initial space-like hypersurface
$\Sigma_I{\,}$.  Here, $h_{ij}{\;}(i,j=1,2,3)$ denotes the 
asymptotically-flat intrinsic spatial metric on $\Sigma_I$ or 
$\Sigma_F{\,}$; both hypersurfaces are diffeomorphic to ${\Bbb R}^3$.  
They are separated by a very large time-interval $T{\,}$, as measured 
at spatial infinity.  The initial boundary data may be chosen (say) 
to be spherically-symmetric, corresponding to a nearly-spherical 
configuration prior to gravitational collapse.  The final data may be 
chosen to register the accumulated spin-0 and spin-2 quantum
radiation, following gravitational collapse, provided that both 
initial and final data have the same mass $M{\,}$, measured at spatial 
infinity.  In order to make the quantum amplitude well-defined, 
Feynman's $+i\epsilon$ approach is taken.  Here, this involves a 
rotation into the complex: $T\rightarrow{\mid}T{\mid}\exp(-i\theta)$, 
with $0<\theta\leq\pi/2{\,}$.  In this case, a complex solution of 
the {\it classical} boundary-value problem is expected to exist, 
whereas, for Lorentzian signature $(\theta =0)$, the classical 
boundary-value problem is badly posed.  For a 
locally-supersymmetric theory, the quantum amplitude should be 
proportional to $\exp(iS_{\rm class})$, apart from corrections which 
are very small when the frequencies in the boundary data are small 
compared to the Planck scale.  Here, $S_{\rm class}$ is the action 
of the classical solution.  The desired Lorentzian amplitude is then 
found by taking the limit $\theta\rightarrow 0_{+}{\,}$.  A Maxwell 
field may be included additionally; it is natural to regard its 
magnetic field $B_i$ as giving additional boundary data on 
$\Sigma_I$ and $\Sigma_F{\,}$.  By a process which parallels exactly 
our previous spin-0-amplitude calculation, one can obtain the 
quantum amplitude for photon data on $\Sigma_F{\,}$.  The magnetic 
boundary conditions are related by supersymmetry to the natural 
spin-2 (gravitational-wave) boundary conditions, which involve fixing 
the magnetic part of the Weyl tensor.
\end{abstract}

\begin{section}{Introduction}
Previous work on this project has been concerned with quantum
amplitudes for scalar fields at late times, following a 
nearly-spherically-symmetric Einstein/massless-scalar collapse to a 
black hole [1-7].  For simplicity, we have assumed that, on an initial 
space-like hypersurface $\Sigma_I$ at some early time, 
spherically-symmetric (Dirichlet) boundary data are specified for the 
gravitational and scalar fields.  After an extremely large
time-interval $T{\,}$, as measured at spatial infinity, a final 
space-like hypersurface $\Sigma_F$ carries final Dirichlet data for 
gravity and the scalar field.  Again for simplicity, we have so far 
considered the case in which the final gravitational data on 
$\Sigma_F$ are also spherically-symmetric, whereas the final scalar 
data include a small non-spherical part $\phi^{(1)}{\,}$.  The 
time-interval $T$ should preferably be taken sufficiently large that 
the evaporation of the black hole is complete before time $T{\,}$.

The quantum amplitude is calculated by means of Feynman's
$+i\epsilon$ approach [8]: one begins by rotating $T$ into the
complex: $T\rightarrow{\mid}T{\mid}\exp(-i\theta)$, where
$0<\theta\leq\pi/2{\,}$.  Then the {\it classical} boundary-value 
problem for the linearised scalar perturbation of the background 
spherically-symmetric Einstein/scalar solution is expected to be
well posed, having a classical Lorentzian action $S_{\rm class}{\,}$. 
(For the Lorentzian-signature case with $\theta =0{\,}$, the 
boundary-value problem is badly posed, being in effect a hyperbolic 
or wave-like boundary-value problem [9-11]).  The Lorentzian quantum 
amplitude for the final scalar configuration $\phi^{(1)}$ is then 
found by applying the limit $\theta\rightarrow 0_{+}$ to the 
semi-classical amplitude, which is proportional to 
$\exp(iS_{\rm class})$, for a locally-supersymmetric theory 
containing Einstein gravity and a massless scalar, as in [9,12], 
apart from loop corrections which should be negligible for boundary 
data involving frequencies below the Planck scale.

This procedure was carried out in detail in [5], for the case of
a real scalar field.  In the present paper, the analogous approach
is applied to the (spin-1) Maxwell field.  That is, the extra
ingredient of a weak (linearised) Maxwell field is combined with
a typical nearly-spherical Einstein/massless-scalar classical
solution, as described above.  In particular, we again study initial 
data on $\Sigma_I$ which are spherically symmetric, containing only 
background gravitational and scalar components.  On the final surface 
$\Sigma_F{\,}$, for simplicity we now take both the gravitational and 
scalar configurations to be exactly spherically symmetric, but include
final data for a weak Maxwell field.  The calculation leads in Sec.5
to an expression for the semi-classical amplitude, proportional to
$\exp(iS^{EM}_{\rm class}){\,}$, as a functional of the final Maxwell
boundary data; here, $S^{EM}_{\rm class}$ denotes the Maxwell
contribution to the classical action.  The resulting Lorentzian quantum 
amplitude is Gaussian in form, for weak Maxwell fields -- just as the 
amplitude of [5] for weak non-spherical final scalar data is Gaussian.

In Sec.2 we consider the Maxwell action and suitable boundary
conditions.  In Sec.3, we summarise the Regge-Wheeler formalism [13] 
for spin-1 (Maxwell) perturbations ofspherically-symmetric black 
holes.  Using this language, we treat the classical Maxwell field 
equations in Sec.4.  The question of appropriate boundary conditions 
is resolved in Sec.5, and the classical Maxwell action 
$S^{EM}_{\rm class}$ is presented as a functional of the final 
spin-1 boundary data.  As mentioned in the previous paragraph, the 
treatment then continues along the same lines as in [5] for spin-0
(massless-scalar) perturbations on the final hypersurface
$\Sigma_F{\,}$.

It is found in Sec.5 that the appropriate boundary data, both for
odd- and even-parity spin-1 perturbations, involve fixing the 
magnetic field on the boundary.  In [14], which also treats the more
complicated spin-2 (gravitational-wave) analogue of the present spin-1
calculation, the treatment again parallels that of the present paper, 
to arrive at a Gaussian expression for the quantum amplitude for 
weak gravitational-wave data on the final hypersurface $\Sigma_F{\,}$.  
In the spin-2 case, the natural boundary conditions for both odd- and 
even-parity perturbations turn out to involve fixing the magnetic part 
of the Weyl tensor [15-18].  It should not be surprising that a
pattern emerges for the appropriate boundary conditions for
different bosonic spins $s=0,1,2{\,}$.  (Had we taken a complex scalar 
field $\phi =\phi_{1}+i\phi_{2}$ in [4-6], such as appears in 
locally-supersymmetric models [19-22], we should have found that, 
while Dirichlet conditions are appropriate for $\phi_{1}{\,}$,
Neumann conditions are needed for $\phi_2{\,}$.)  In Teukolsky's
essentially spinor-based approach [23,24], which developed from 
around 1973, the classical field equations for spins 
$s=0,{1\over 2},1,{3\over 2}{\,}$ and $2{\,}$, perturbed around a 
Kerr black-hole background, all fit into an s-dependent pattern.  
This question is clearly intimately connected with that of local 
supersymmetry; the Dirichlet/Neumann division for a complex scalar 
field arose in 1982 in the context of boundary conditions for gauged 
supergravity [20-22].  As part of a more unified description, the 
spin-1 magnetic boundary conditions will be described in the language 
of 2-component spinors [9,17,18] in Sec.6.  The corresponding 
2-spinor description of magnetic Weyl-tensor boundary conditions is 
also discussed briefly in Sec.7 of [14].  Our work on 
spin-${{1}\over{2}}$ amplitudes is described in [25].  The remaining 
fermionic case of spin-${{3}\over{2}}$ amplitudes is in 
preparation [26]; this is needed for a further understanding of 
local supersymmetry in the present context.
\end{section}

\begin{section}{The Maxwell Equations}
The Maxwell contribution to the total Lorentzian action $S$ is
$$S^{EM}{\;}{\,} 
={\;}{\,}-{\,}{1 \over{16\pi}}{\;}\int_{{\cal M}}{\;}d^{4}x{\;}{\,}
(-g)^{1\over2}{\;}F_{\mu\nu}{\,}F^{\mu\nu}{\quad},\eqno(2.1)$$
\noindent
where $F_{\mu\nu}=F_{[\mu\nu]}$ is the Maxwell field strength, while
$g_{\mu\nu}$ is the space-time metric, assumed here to have Lorentzian
signature, with $g={\rm det}(g_{\mu\nu})<0{\,}.$  The resulting 
classical Maxwell field equations are
$$\nabla_{\mu}{\,}F^{\mu\nu}{\;}{\,}={\;}{\,}0{\quad}.\eqno(2.2)$$

The further condition that $F_{\mu\nu}$ be derivable from a vector
potential $A_{\mu}{\,}$, as
$$F_{\mu\nu}{\;}{\,}
={\;}{\,}\nabla_{\mu}A_{\nu}{\,}-{\,}\nabla_{\nu}A_{\mu}{\;}{\,}
={\;}{\,}\partial_{\mu}A_{\nu}{\,}-{\,}\partial_{\nu}A_{\mu}{\quad},
\eqno(2.3)$$  
\noindent
may equivalently be written in the form of the dual field equations
$$\nabla_{\mu}{\,}(^{*}F^{\mu\nu}){\;}{\,}={\;}{\,}0{\quad},\eqno(2.4)$$
\noindent
where
$$^{*}F_{\mu\nu}{\;}{\,}={\;}{\,}{1\over 2}{\;}
\eta_{\alpha\beta\mu\nu}{\;}F^{\alpha\beta}\eqno(2.5)$$ 
\noindent
is the dual field strength [17,18].  Here, 
${\,}\eta_{\alpha\beta\mu\nu}{\,}$
is the totally-antisymmetric tensor with components [27] 
(in Lorentzian signature) 
$$\eta_{\alpha\beta\mu\nu}{\;}{\,}={\;}{\,}\bigl(-g\bigr)^{1/2}{\;}
\epsilon_{\alpha\beta\mu\nu}{\quad},\eqno(2.6)$$
where ${\,}g{\,}={\,}{\rm det}(g_{\mu\nu}){\,}$ and
${\,}\epsilon_{\alpha\beta\mu\nu}$ is the alternating symbol in 
4 dimensions.  (Note here that the ${\rm Poincar\acute e}$ lemma [28]  
may be applied, since we are working within a manifold ${\cal M}$ 
which may be regarded as a slice of ${\Bbb R}^4$, with boundary 
$\partial{\cal M}$ consisting of two ${\Bbb R}^3$ hypersurfaces.)  
The action (2.1) is invariant under Maxwell gauge transformations
$$A_{\mu}\rightarrow{\;}A_{\mu}{\;}+{\;}\partial_{\mu}\Lambda{\quad},
\eqno(2.7)$$ 
\noindent
in the interior, where $\Lambda(x)$ is a function of position.

As in [4-6] for scalar (spin-0) perturbations of spherical 
Einstein/massless-scalar gravitational collapse, and as in the 
treatment of spin-2 (graviton) perturbations, we shall need the 
classical action $S_{\rm class}{\,}$, namely the action $S$ evaluated 
at a classical solution of the appropriate (slightly complexified) 
boundary-value problem, in order to obtain the semi-classical quantum 
amplitude, proportional to $\exp(iS_{\rm class})$, and hence by a 
limiting procedure to obtain the Lorentzian quantum amplitude.  
In the present (spin-1) Maxwell case, the classical action 
$S^{EM}_{\rm class}$ resides solely on the boundary $\partial M$, 
which consists of the initial space-like hypersurface $\Sigma_I$ and 
final hypersurface $\Sigma_F{\,}$.  There will be no contribution 
from any large cylinder of radius $R_{\infty}\rightarrow\infty{\,}$, 
provided we impose the physically reasonable restriction that the 
potential $A_{\mu}$ die off faster than $r^{-1}$, and the field 
strength $F_{\mu\nu}$ faster than $r^{-2}$, as 
$r\rightarrow\infty{\,}$.  That is, we impose reasonable fall-off 
conditions at large $r$ on field configurations, such that the action 
$S$ should be finite.  (Compare the usual fall-off conditions for 
instantons in Euclidean Yang-Mills theory [28-30].)  For the above
class of Maxwell field configurations, the boundary form of the
classical Maxwell action is
$$S^{EM}_{~~~ \rm class}{\;}{\,}
={\;}{\,}-{\,}{1 \over{8\pi}}{\,}\int^{\Sigma_F}_{\Sigma_I}{\;} 
d^{3}x{\;}{\,}h^{1\over 2}{\;}n_{\mu}{\;}A_{\nu}{\;}F^{\mu\nu}{\quad}.
\eqno(2.8)$$
\noindent
Here $h_{ij}=g_{ij}{\;}(i,j=1,2,3)$ gives the intrinsic Riemannian 
3-metric on the boundary hypersurface $\Sigma_I$ or $\Sigma_F{\,}$, 
and we write ${\,}h={\rm det}(h_{ij})>0{\,}$.  Further, $n^{\mu}$
denotes the (Lorentzian) unit timelike vector, normal to the
space-like hypersurface $\Sigma_I$ or $\Sigma_F{\,}$.

Given the ${\,}(3{\,}+{\,}1)$ split of the 4-metric $g_{\mu\nu}$ at each
boundary, due to the ability to project vectors and tensors normally
using $n^{\mu}$ and tangentially to $\Sigma_I$ or $\Sigma_F$ using the
projector [27]
$$h_{\mu\nu}{\;}{\,}
={\;}{\,}g_{\mu\nu}{\;}+{\;}n_{\mu}{\,}n_{\nu}{\quad},\eqno(2.9)$$
\noindent
at the boundary, one can project the potential $A_{\mu}$ and field
strength $F_{\mu\nu}$ into 'normal' and 'spatial' parts on $\Sigma_I$
and $\Sigma_F{\,}$.  In particular, one defines the densitised electric
field vector on the boundary:
$${\cal E}^{i}{\;}{\,}
={\;}{\,}-{\,}h^{1\over 2}{\;}E^{i}{\quad},\eqno(2.10)$$
\noindent
where
$$E^{\nu}{\;}{\,}={\;}{\,}n_{\mu}{\,}F^{\mu\nu}{\quad}.\eqno(2.11)$$
\noindent
Further, in a Hamiltonian formulation [31], when one regards the
spatial components $A_i$ of the vector potential as 'coordinates',
the canonical momentum $\pi^i{\,}$, automatically a vector density,
is given by
$$\pi^{i}{\;}{\,}={\;}{\,}-{\,}{{{\cal E}^{i}}\over{4\pi}}{\quad}.
\eqno(2.12)$$
\noindent
Note that the normal component $A_{t}=-\varphi{\,}$, where $\varphi$ is
the scalar potential, is gauge-dependent, but that $\varphi$ does not
need to be specified on the spacelike boundaries $\Sigma_I$ and
$\Sigma_F{\,}$, and is indeed allowed to vary freely there and throughout
the space-time.  Its conjugate momentum therefore vanishes.  In the
gravitational case, analogous properties hold for the lapse function 
$N$ and shift vector $N^i$ [9,31].

As described in [31-33], it is most natural in specifying a classical
boundary-value problem for the Maxwell field, with data given on the
space-like boundaries $\Sigma_I$ and $\Sigma_F{\,}$, together with a
Lorentzian time-separation $T{\,}$, as measured at spatial infinity, 
to fix the spatial magnetic field components, described in densitised
form by 
$${\cal B}^{i}{\;}{\,}
={\;}{\,}{1\over 2}{\;}\epsilon^{ijk}{\;}F_{jk}{\quad},\eqno(2.13)$$
\noindent
on $\Sigma_I$ and $\Sigma_F{\,}$.  The ${\cal B}^i$ cannot be specified
freely on the boundary, but only subject to the restriction
$$\partial_{i}{\cal B}^{i}{\;}{\,}={\;}{\,}0{\quad}.\eqno(2.14)$$
\noindent
These components are gauge-invariant, and therefore physically
measurable, in contrast to those of the spatially-projected vector
potential $A_i{\,}$.  We shall regard the space of such 
${\cal B}^{i}(x){\,}$, on $\Sigma_I$ or $\Sigma_F{\,}$, as the 
'coordinates' for Maxwell theory.  From the space-time Maxwell 
equations (2.2), one also deduces the constraint
$$\partial_{i}{\cal E}^{i}{\;}{\,}={\;}{\,}0{\quad}.\eqno(2.15)$$
\end{section}

\begin{section}{The Regge-Wheeler formalism}
In 1957, Regge and Wheeler [13] developed the formalism for treating
both spin-1 and spin-2 classical perturbations of the Schwarzschild
solution and of other spherically-symmetric solutions, corresponding
to Maxwell and gravitational (graviton) perturbations.  Here, in the
Regge-Wheeler (RW) formalism, for Maxwell theory we decompose the real
linearised field strength $F^{(1)}_{\mu\nu}$ and linearised vector
potential $A^{(1)}_{\mu}$ into tensor and vector spherical harmonics,
respectively [34,35].  We are assuming that the background 
(unperturbed) classical solution consists, as in [4-7], of a 
spherically-symmetric gravitational and massless-scalar field 
$(\gamma_{\mu\nu}{\,},\Phi)$, with no background Maxwell field:  
$A^{(0)}_{\mu}=0{\,},{\;}F^{(0)}_{\mu\nu}=0{\,}$.

For each spin $s=0,1,2{\,}$, the corresponding perturbation modes
split into those with even parity and those with odd parity.  Under
the parity inversion: 
$\theta\rightarrow(\pi -\theta){\,},{\,}\phi\rightarrow(\pi +\phi){\,}$, 
we define the even perturbations as those with parity 
$\pi=(-1)^{\ell}{\,}$, while the odd perturbations have parity 
$\pi=(-1)^{\ell +1}{\,}$.  For Maxwell theory $(s=1)$, the $\ell=0$ 
mode corresponds to a static perturbation, in which a small amount of 
electric charge is added to the black hole; in particular, 
a Schwarzschild solution will be 'displaced' infinitesimally along 
the family of Reissner-Nordstr\"om solutions.  For radiative modes 
with $\ell =1$ (dipole) and higher, we set
$$\eqalignno{F^{(1)}_{\mu\nu}(x){\;}{\,}&
={\;}{\,}\sum^{\infty}_{\ell=1}{\;}\sum^{\ell}_{m=-\ell}{\;}{\,} 
\biggl[\Bigl(F^{(o)}_{\mu\nu}\Bigr)_{\ell m}{\,}
+{\;}\Bigl(F^{(e)}_{\mu\nu}\Bigr)_{\ell m}\biggr]{\quad},&(3.1)\cr
A^{(1)}_{\mu}(x){\;}{\,}&
={\;}{\,}\sum^{\infty}_{\ell=1}{\;}\sum^{\ell}_{m=-\ell}{\;}{\,}
\biggl[\Bigl(A^{(o)}_{\mu}\Bigr)_{\ell m}{\,}
+{\;}\Bigl(A^{(e)}_{\mu}\Bigr)_{\ell m}\biggr]{\quad}.&(3.2)\cr}$$
\noindent
On substituting this decomposition into the boundary expression (2.7)
for the classical Maxwell action $S^{EM}_{~~~ \rm class}{\,}$, we find
$$\eqalign{S^{EM}_{\rm class}{\,}
={\,}&-{\,}{1\over{8\pi}}\sum_{\ell m\ell' m'}
\int d\Omega\int^{R_{\infty}}_{0}dr{\;}r^{2}{\,}e^{(a-b)/2}{\;}
\gamma^{ij}{\,}\Bigl(A^{(o)}_{j}\Bigr)_{\ell m}{\,}
\Bigl(F^{(o)}_{ti}\Bigr)^{*}_{\ell' m'}
\Bigl\bracevert^{\Sigma_F}_{\Sigma_I}\cr
&-{\,}{1\over{8\pi}}\sum_{\ell m\ell' m'}\int d\Omega
\int^{R\infty}_{0}dr{\;}r^{2}{\,}e^{(a-b)/2}{\;}
\gamma^{ij}{\,}\Bigl(A^{(e)}_{j}\Bigr)_{\ell m}{\,}
\Bigl(F^{(e)}_{ti}\Bigr)^{*}_{\ell' m'}
\Bigr\bracevert^{\Sigma_F}_{\Sigma_I}{\;},\cr}\eqno(3.3)$$  
\noindent
where $\int d\Omega$ denotes integration over the sphere, with respect 
to the angular coordinates $\theta{\,},\phi{\,}$, and the 
spherically-symmetric background metric is taken in the form [4-6]:
$$ds^{2}{\;}{\,}
={\;}-{\,}e^{b(t,r)}{\,}dt^{2}+{\,}e^{a(t,r)}{\,}dr^{2} 
+{\,}r^{2}{\,}(d\theta^{2}+{\,}\sin^{2}\theta{\;}d\phi^{2}){\quad}.
\eqno(3.4)$$
\noindent
For later reference, we define $m(t,r)$ by
$$e^{-a}{\;}{\,}={\;}{\,}1{\,}-{\,}{{2m(t,r)}\over r}{\quad},\eqno(3.5)$$
\noindent
within the region of the space-time where the black hole is
evaporating.  Clearly, the odd and even contributions decouple. 

For the subsequent detailed treatment of the angular harmonics involved,
we follow Zerilli's decomposition [36] of $F^{(1)}_{\mu\nu}$ and
$A^{(1)}_{\mu}{\,}$.  We set
$$F^{(1)}_{\mu\nu}(x){\;}{\,}
={\;}{\,}\sum_{\ell m}{\;}\Bigl(F^{(1)}_{\mu\nu}\Bigr)_{\ell m}{\quad},
\eqno(3.6)$$ 
$$\Bigl(F^{(1)}_{\mu\nu}\Bigr)_{\ell m}{\;}{\,}
={\;}{\,}\pmatrix{0& {f_{1}}& {f_{2}}& {f_{3}}\cr
{-f_{1}}& 0& {f_{4}}&{f_{5}}&\cr
{-f_{2}}& {-f_{4}}& 0& {f_{6}}\cr
{-f_{3}}& {-f_{5}}& {-f_{6}}& 0\cr}\eqno(3.7)$$
\noindent
For a given choice of $(\ell,m)$, we take
$$\eqalignno{f_{1}{\;}{\,}&
={\;}{\,}\bigl(\hat F^{(e)}_{tr}\bigr)_{\ell m}{\;}
Y_{\ell m}(\Omega){\quad},&(3.8)\cr
f_{2}{\;}{\,}&
={\;}{\,}\bigl(\sin\theta\bigr)^{-1}{\;}
\bigl(\hat F^{(o)}_{t\theta}\bigr)_{\ell m}{\;}
\bigl(\partial_{\phi}Y_{\ell m}\bigr){\;}
+{\;}\bigl(\hat F^{(e)}_{t\theta}\bigr)_{\ell m}{\;}
\bigl(\partial_{\theta}Y_{\ell m}\bigr){\quad},&(3.9)\cr
f_{3}{\;}{\,}&
={\;}{\,}-{\,}\bigl(\hat F^{(o)}_{t\theta}\bigr)_{\ell m}{\,}
\bigl(\sin\theta\bigr){\;}\bigl(\partial_{\theta}Y_{\ell m}\bigr){\;}
+{\,}\bigl(\hat F^{(e)}_{t\theta}\bigr)_{\ell m}{\;}
\bigl(\partial_{\phi}Y_{\ell m}\bigr){\quad},&(3.10)\cr
f_{4}{\;}{\,}&
={\;}{\,}\bigl(\sin\theta\bigr)^{-1}{\;}
\bigl(\hat F^{(o)}_{r\theta}\bigr)_{\ell m}{\;}
\bigl(\partial_{\phi}Y_{\ell m}\bigr){\;} 
+{\;}\bigl(\hat F^{(e)}_{r\theta}\bigr)_{\ell m}{\;}
\bigl(\partial_{\theta}Y_{\ell m}\bigr){\quad},&(3.11)\cr
f_{5}{\;}{\,}&
={\;}{\,}-{\,}\bigl(\hat F^{(o)}_{r\theta}\bigr)_{\ell m}
\bigl(\sin\theta\bigr){\;}\bigl(\partial_{\theta}Y_{\ell m}\bigr){\;}
+{\;}\bigl(\hat F^{(e)}_{r\theta}\bigr)_{\ell m}{\,}
\bigl(\partial_{\phi}Y_{\ell m}\bigr){\quad},&(3.12)\cr
f_{6}{\;}{\,}&
={\;}{\,}\bigl(\hat F^{(o)}_{\theta\phi}\bigr)_{\ell m}{\;}
\bigl(\sin\theta\bigr){\;}Y_{\ell m}(\Omega){\quad}.&(3.13)\cr}$$
\noindent
Here, the $Y_{\ell m}(\Omega)$ are scalar spherical harmonics [37],
and a caret indicates that the quantity is a function of $t$ and
$r$ only.

Again, following [13], for the vector potential, we set
$$\eqalignno{\Bigl(A^{(o)}_{\mu}\Bigr)_{\ell m}(x) 
&=\biggl(0,0,
{{a_{2\ell m}(t,r)(\partial_{\phi}Y_{\ell m})}
\over{\ell(\ell +1)\sin\theta}},
{{-a_{2\ell m}(t,r)(\sin\theta)
(\partial_{\theta}Y_{\ell m})}\over{\ell(\ell +1)}}\biggr),
&(3.14)\cr
\Bigl(A^{(e)}_{\mu}\Bigr)_{\ell m}(x)
&=\biggl(-{\,}a_{0\ell m}(t,r){\,}Y_{\ell m}(\Omega){\,},{\,}
a_{1\ell m}(t,r){\,}Y_{\ell m}(\Omega){\,},{\,}0{\,},{\,}0\biggr).
&(3.15)\cr}$$
\noindent
Eq.(3.3) can now be expanded out in the form:
$$\eqalign{S^{EM}_{\rm class}{\;}
={\,}&-{\,}{{1}\over{8\pi}}\sum_{\ell m\ell' m'}{\,}\int d\Omega{\,}
\int^{R_{\infty}}_{0}{\,}dr{\;}{\,}e^{(a-b)/2}{\quad}\times\cr
&{\qquad}\times
\biggl[\Bigl(A_{\theta}^{(o)}\Bigr)_{\ell m}
\Bigl(F^{(o)}_{t\theta}\Bigr)^{*}_{\ell' m'} 
+(\sin\theta)^{-2}{\,}\Bigl(A^{(o)}_{\phi}\Bigr)_{\ell m}
\Bigl(F^{(o)}_{t\phi}\Bigr)^{*}_{\ell' m'}\biggr]
\Bigl\bracevert^{\Sigma_F}_{\Sigma_I}\cr
&-{\,}{{1}\over{8\pi}}\sum_{\ell m \ell' m'}\int d\Omega
\int^{R_{\infty}}_{0}dr{\;}r^{2}{\,}e^{(a-b)/2}{\,}
\Bigl(A^{(e)}_{r}\Bigr)_{\ell m}\Bigl(F^{(e)}_{t,r}\Bigr)^{*}_{\ell' m'}
\Bigl\bracevert^{\Sigma_F}_{\Sigma_I}{\,}.\cr}\eqno(3.16)$$

Of course, the components of the field strength are given in terms of
those of the vector potential by Eq.(2.3); for example,
$F^{(1)}_{t\theta}
=(\partial_{t}A^{(1)}_{\theta}-\partial_{\theta}A^{(1)}_{t}).$
This gives the relations
$$\eqalignno{\Bigl(\hat F^{(o)}_{t\theta}\Bigr)_{\ell m}{\;}{\,}
&={\;}{\,}{{\bigl(\partial_{t}a_{2\ell m}\bigr)}
\over{\ell(\ell +1)}}{\quad},&(3.17)\cr
\Bigl(\hat F^{(e)}_{t\theta}\Bigr)_{\ell m}{\;}{\,}
&={\;}{\,}a_{0\ell m}{\quad},&(3.18)\cr
\Bigl(\hat F^{(o)}_{r\theta}\Bigr)_{\ell m}{\;}{\,}
&={\;}{\,}{{\bigl(\partial_{r}a_{2\ell m}\bigr)
\over{\ell(\ell+1)}}}{\quad},&(3.19)\cr
\Bigl(\hat F^{(e)}_{r\theta}\Bigr)_{\ell m}{\;}{\,}
&={\;}{\,}a_{1\ell m}{\quad},&(3.20)\cr
\Bigl(\hat F^{(e)}_{tr}\Bigr)_{\ell m}{\;}{\,}
&={\;}{\,}\bigl(\partial_{r}a_{0\ell m}\bigr){\;}
-{\;}\bigl(\partial_{t}a_{1\ell m}\bigr){\quad},&(3.21)\cr
\Bigl(\hat F^{(o)}_{\theta\phi}\Bigr)_{\ell m}{\;}{\,}
&={\;}{\,}a_{2\ell m}{\quad}.&(3.22)\cr}$$
\noindent
The action (3.16) then simplifies to give
$$\eqalign{S^{EM}_{\rm class}{\;} 
={\,}&-{\,}{{1}\over{8\pi}}{\,}\sum_{\ell m}{\;}\ell(\ell +1){\;}
{\int}^{R_{\infty}}_{0}{\,}dr{\;}{\,}e^{(a-b)/2}{\;}f^{(o)}_{\ell m}{\;}
\Bigl(\partial_{t}f^{(o)*}_{\ell m}\Bigr)
\Bigl\bracevert^{\Sigma_F}_{\Sigma_I}\cr
&-{\,}{{1}\over{8\pi}}{\,}\sum_{\ell m}{\,}{\int}^{R_{\infty}}_{0}
dr{\;}r^{2}{\,}e^{-(a+b)/2}{\,}a_{1\ell m}
\Bigl(\bigl(\partial_{t}a^{*}_{1\ell m}\bigr)
-\bigl(\partial_{r}a^{*}_{0\ell m}\bigr)\Bigr)
\Bigl\bracevert^{\Sigma_F}_{\Sigma_I}{\,},\cr}\eqno(3.23)$$
\noindent
where
$$f^{(o)}_{\ell m}{\;}{\,}
={\;}{\,}{{a_{2\ell m}}\over{\ell(\ell +1)}}\eqno(3.24)$$
\noindent
determines the odd-parity Maxwell tensor {\it via} Eqs.(3.17,19,22).
\end{section}

\begin{section}{Field equations}
The form of the classical action (3.22) can be further simplified by
using the Maxwell field equations (2.2,4).  This will lead finally to
the form (4.14) below, in which $S^{EM}_{\rm class}$ is expressed
explicitly in terms of boundary data, as needed in the subsequent
calculation of the quantum amplitude (see [4-6] for the spin-0
analogue).

The linearised Maxwell equations can be written as
$$F^{(1)\mu\nu}_{~~~~~~;\nu}{\;}{\,}
={\;}{\,}(-\gamma)^{-{{1}\over{2}}}{\;}
\partial_{\nu}\bigl((-\gamma)^{1\over2}{\,}F^{(1)\mu\nu}\bigr){\;}{\,}
={\;}{\,}0{\quad}.\eqno(4.1)$$
\noindent
The ${\,}\mu =t,r{\,}$ equations give
$$\eqalignno{\partial_{r}
\bigl(r^{2}\bigl(\hat F^{(e)}_{tr}\bigr)_{\ell m}\bigr){\,}
-{\,}\ell(\ell+1){\;}e^{a}{\;}
\bigl(\hat F^{(e)}_{t\theta}\bigr)_{\ell m}{\;}{\,}
&={\;}{\,}0{\quad},&(4.2)\cr
e^{a}{\;}\partial_{t}\bigl(\hat F^{(e)}_{t\theta}\bigr)_{\ell m} 
-{\;}\partial_{r}\bigl(e^{-a}{\,}
\bigl(\hat F^{(e)}_{r\theta}\bigr)_{\ell m}\bigr){\;}{\,}
&={\;}{\,}0{\quad},&(4.3)\cr 
\partial_{r}\bigl(e^{-a}{\,}
\bigl(\hat F^{(o)}_{r\theta}\bigr)_{\ell m}\bigr){\,}
-{\;}e^{a}{\;}\partial_{t}\bigl(\hat F^{(o)}_{t\theta}\bigr)_{\ell m}{\,} 
-{\,}r^{-2}{\;}\bigl(\hat F^{(o)}_{\theta\phi}\bigr)_{\ell m}{\;}{\,}
&={\;}{\,}0{\quad},&(4.4)\cr
\partial_{t}\bigl(\hat F^{(e)}_{tr}\bigr)_{\ell m}{\,} 
-{\;}r^{-2}{\;}e^{-a}{\;}\ell(\ell +1){\;}
\bigl(\hat F^{(e)}_{r\theta}\bigr)_{\ell m}{\;}{\,}
&={\;}{\,}0{\quad}.&(4.5)\cr}$$
\noindent
The ${\,}\mu=\theta ,\phi{\,}$ components give the same equations.  
Note that Eq.(4.2) is just the (source-free) constraint equation
${\,}\partial_{i}{\cal E}^{(1)i}{\,}={\,}0{\,}$ of Eq.(2.14).

The equations (3.17,19,22) together imply the decoupled wave
equation for odd perturbations:
$$\bigl(\partial_{r^*}\bigr)^{2}{\,}a_{2\ell m}{\,}
-{\,}\bigl(\partial_{t}\bigr)^{2}{\,}a_{2\ell m}{\,}
-{\,}V_{1\ell}(r){\;}a_{2\ell m}{\;}{\,} 
={\;}{\,}0{\quad},\eqno(4.6)$$ 
\noindent
where
$$V_{1\ell}(r){\;}{\,}
={\;}{\,}{{e^{-a}{\;}\ell(\ell+1)}\over{r^{2}}}{\quad}
>{\quad}0\eqno(4.7)$$
\noindent
is the (massless) spin-1 effective potential and where, as usual, we
write ${\,}\partial_{r^*}=e^{-a}{\,}\partial_r{\,}.$  As in [4-7],
we assume that the adiabatic approximation is valid in a
neighbourhood of the initial and final surfaces, $\Sigma_I$ and
$\Sigma_F{\,}$.  In that case, we can, as before, effectively work with
the field equations on a Schwarzschild background, except that the
Schwarzschild mass $M_{0}$ is replaced by a mass function $m(t,r)$,
as in Eq.(3.5), which varies extremely slowly with respect both 
to time and to radius.

Equation (4.3) gives 
$\partial_{t}\bigl(\hat F^{(e)}_{t\theta}\bigr)_{\ell m}$ 
in terms of $\bigl(\hat F^{(e)}_{r\theta}\bigr)_{\ell m}{\,}$, 
while Eq.(4.5) gives
${\,}\partial_{t}\bigl(\hat F^{(e)}_{tr}\bigr)_{\ell m}$ in terms of 
$\bigl(\hat F^{(e)}_{r\theta}\bigr)_{\ell m}{\,}$.  Together, 
Eqs.(3.21, 4.3,5) imply that
$${\bigr(\partial _{r^*}\bigr)}^{2}f^{(e)}_{\ell m}{\,}
-{\,}\bigl({\partial}_{t}\bigr)^{2}f^{(e)}_{\ell m}{\,}
-{\,}V_{1\ell}{\;}f^{(e)}_{\ell m}{\;}{\,} 
={\;}{\,}0{\quad},\eqno(4.8)$$
where we define
$$f^{(e)}_{\ell m}{\;}{\,}={\;}{\,}e^{-a}{\;}a_{1\ell m}{\quad}.\eqno(4.9)$$
\noindent
Thus, with a suitably defined variable ${\,}f^{(e)}_{\ell m}{\,}$, 
the even perturbations obey the same decoupled wave equation (4.6) 
as the odd perturbations. 

Finally [38], we set
$$\psi^{(e)}_{\ell m}(t,r){\;}{\,}
={\;}{\,}{{r^{2}}\over{\ell(\ell+1)}}{\;}
\Bigl(\bigl(\partial_{t}a_{1\ell m}\bigr){\,}
-{\,}\bigl(\partial_{r}a_{0\ell m}\bigr)\Bigr){\quad},\eqno(4.10)$$
\noindent
which is clearly gauge-invariant.  Now, $\psi^{(e)}_{\ell m}$ 
is simply related to the (even-parity) function $f^{(e)}_{\ell m}$ 
of the previous paragraph: Eqs.(4.3,8) imply that
$$\bigl(\partial_{t}\psi^{(e)}_{\ell m}\bigr){\;}{\,}
={\;}{\,}-{\,}f^{(e)}_{\ell m}{\quad}.\eqno(4.11)$$ 

Equations (4.9,10)  can now be used to simplify the classical Lorentzian
action (3.23).  For ease of comparison with the (second-variation)
classical spin-2 action, where the pattern is similar, we define
$$\eqalignno{\psi^{(o)}_{1\ell m}{\;}{\,}
&={\;}{\,}a_{2\ell m}(t,r){\quad},&(4.12)\cr
\psi^{(e)}_{1\ell m}{\;}{\,}
&={\;}{\,}\ell(\ell+1){\;}\psi^{(e)}_{\ell m}(t,r){\quad}.&(4.13)\cr}$$ 
\noindent
Then, given weak-field Maxwell boundary data specified by the
linearised magnetic-field mode components ${\,}\{B^{(1)i}_{\ell m}\}$ 
on each of the boundaries $\Sigma_I$ and $\Sigma_F{\,}$, 
the corresponding classical Maxwell action is
$$\eqalign{S^{EM}_{\rm class}&\biggl[\Bigl\{B^{(1)i}_{\ell m}\Bigr\}\biggr]\cr 
&={\,}{{1}\over{8\pi}}{\,}\sum_{\ell m}{\,}
{{(\ell-1)!}\over{(\ell + 1)!}}{\;}{\int}^{R_{\infty}}_{0}
dr{\;}e^{a}{\,}\Bigl(\psi^{(e)}_{1\ell m}{\,}
\bigl(\partial_{t}\psi^{(e)*}_{1\ell m}\bigr){\,}
-{\,}\psi^{(o)}_{1\ell m}{\,}
\bigl(\partial_{t}\psi^{(o)*}_{1\ell m}\bigr)\Bigr)
\Bigl\bracevert^{\Sigma_F}_{\Sigma_I}{\,}.\cr}\eqno(4.14)$$
\noindent
Of course, the limit $R_{\infty}\rightarrow\infty{\,}$ must be 
understood in Eq.(4.14).

Note further that, from Eqs.(3.18,4.2), one has
$$\bigl(\partial_{r*}\psi^{(e)}_{\ell m}\bigr){\;}{\,} 
={\;}{\,}-{\,}{\,}a_{0\ell m}{\quad},\eqno(4.15)$$
\noindent
whence $\psi^{(e)}_{1\ell m}$ also obeys the same decoupled wave
equation (4.6,8) as for $a_{2\ell m}$ and for $f^{(e)}_{\ell m}{\,}$,
namely,
$$\bigl(\partial_{r*}\bigr)^{2}{\,}\psi^{(e)}_{1\ell m}{\;} 
-{\,}\bigl(\partial_{t}\bigr)^{2}{\,}\psi^{(e)}_{1\ell m}{\;}
-{\,}V_{1\ell}{\;}\psi^{(e)}_{1\ell m}{\;}{\,}
={\;}{\,}0{\quad}.\eqno(4.16)$$ 
\noindent
The spin-1 radial equation (4.6,8,16) in a Schwarzschild background,
both for odd- and even-parity Maxwell fields, was first given in 1962
by Wheeler [39].

This suggests a 'preferred route' for understanding the even-parity
perturbations (which are more complicated than in the odd-parity case,
which only involves the single function $a_{2\ell m}(t,r)$, obeying
the decoupled field equation (4.6)):  Given suitable boundary
conditions, one first solves the linear decoupled wave equation in two
variables $t$ and $r{\,}$, namely, Eq.(4.16), for 
$\psi^{(e)}_{1\ell m}(t,r)$.  By differentiation, following 
Eqs.(4.13,15), one then finds $a_{0\ell m}(t,r)$.  Then, 
by integrating Eq.(4.10), one obtains also $a_{1\ell m}(t,r)$, 
and hence, from Eq.(4.9), $f^{(e)}_{\ell m}(t,r)$.  
From Eqs.(3.17-22), one has all the non-zero components of the 
Maxwell field strength $F_{\mu\nu}$ in this even-parity case.
\end{section}

\begin{section}{Boundary Conditions}
Physically, our gauge-invariant odd- and even-parity variables
$\psi^{(o)}_{1\ell m}\;,\;\psi^{(e)}_{1\ell m}$ are effectively the radial
components of the magnetic and electric field strengths, respectively:
$$\eqalignno{B^{(1)r}_{\ell m}(x){\;}{\,}
&={\;}{\,}{{e^{-(a/2)}}\over{r^{2}}}{\;}{\,}\psi^{(o)}_{1\ell m}(t,r){\;}{\,} 
Y_{\ell m}(\Omega){\quad},&(5.1)\cr
E^{(1)r}_{\ell m}(x){\;}{\,}
&={\;}{\,}-{\;}{{e^{-(a/2)}}\over{r^{2}}}{\;}{\,}
\psi^{(e)}_{1\ell m}(t,r){\;}{\,}
Y_{\ell m}(\Omega){\quad},&(5.2)\cr}$$
\noindent
where
$$B^{(1)i}_{\ell m}{\;}{\,} 
={\;}{\,}(^{(3)}\gamma)^{-{1\over 2}}{\;}{\,}
{\cal B}^{(1)i}_{\ell m}{\quad}.\eqno(5.3)$$
\noindent
The remaining, transverse, magnetic field components are
$$\eqalignno{B^{(1)\theta}_{\ell m}(x){\,}
&={\,}{{e^{-(a/2)}}\over{r^{2}\bigl(\sin\theta\bigr)}}
\biggl(a_{1\ell m}(t,r)\bigl(\partial_{\phi}Y_{\ell m}\bigr) 
+{{\bigl(\partial_{r}\psi^{(o)}_{1\ell m}\bigr)}\over{\ell(\ell+1)}}
\bigl(\sin\theta\bigr)\bigl(\partial_{\theta}Y_{\ell m}\bigr)\biggr).
&(5.4)\cr
B^{(1)\phi}_{\ell m}(x){\,}
&={\,}{{e^{-(a/2)}}\over{r^{2}\bigl(\sin^{2}\theta\bigr)}}
\biggl(-a_{1\ell m}(t,r)
\bigl(\sin\theta\bigr)\bigl(\partial_{\theta}Y_{\ell m}\bigr)
+{\bigl({\partial_{r}\psi^{(o)}_{1\ell m}\bigr)}\over{\ell(\ell+1)}}
\bigl(\partial_{\phi}Y_{\ell m}\bigr)\biggr).&(5.5)\cr}$$
\ss
\indent
The main aim of this paper is to calculate quantum amplitudes for weak
spin-1 (Maxwell) perturbative data on the late-time final surface 
$\Sigma_F{\,}$, by evaluating the classical action 
$S^{EM}_{\rm class}$  and hence the semi-classical wave function 
${\rm (const.)}{\,}\times{\,}\exp(iS^{EM}_{\rm class})$, 
{\it as a functional of the spin-1 final boundary data}.  
In Eq.(4.14), ${\,}S^{EM}_{\rm class}$ was expressed as an integral 
over the boundary, involving various perturbative quantities used in 
the description above of the dynamical perturbations.  The present
task is to determine 'optimal' or 'natural' boundary data, both for 
the odd-parity case and separately for the even-parity case, such that 
(i.) the classical boundary-value problem can readily be solved, given 
these data, and (ii.) the classical Maxwell action 
$S^{EM}_{\rm class}$ can be (re-)expressed in terms of the appropriate 
boundary data.  Under those conditions, we will then have a
description of the spin-1 radiation, associated with gravitational 
collapse to a black hole, analogous to that for the spin-0 
(massless-scalar) radiation, as developed in [4,5]. 

Following the discussion of Secs.2 and 4, the relevant field
components to be fixed on $\Sigma_I$ and $\Sigma_F$ are, in the
odd-parity case, $\psi^{(o)}_{1\ell m}{\,}$, as may be seen from 
Eq.(5.1).  In the even-parity case, it is relevant to fix 
$a_{1\ell m}{\,}$, as may be seen from Eqs.(5.4,5).  For the even 
case, from Eqs.(4.9,11), this is equivalent to specification of 
$\bigl(\partial_{t}\psi^{(e)}_{1\ell m}\bigr)$ on the boundary.  
Thus, we choose the boundary data to consist of
$\psi^{(o)}_{1\ell m}$ in the odd-parity case, and
$\bigl(\partial_{t}\psi^{(e)}_{1\ell m}\bigr)$ in the even-parity
case.  Hence, even though $\psi^{(o)}_{1\ell m}$ and 
$\psi^{(e)}_{1\ell m}$ both obey the same dynamical field equations 
(4.6,12) and (4.16), the natural boundary conditions for them on 
$\Sigma_{I,F}$ are quite different -- Dirichlet for the odd-parity 
case $\psi^{(o)}_{1\ell m}{\,}$, but Neumann in the even-parity case 
$\psi^{(e)}_{1\ell m}{\,}$.  This is reminiscent of the situation 
obtaining when spin-2 gravity is coupled to all lower spins, fermionic 
as well as bosonic; that is, to spins 
$s={{3}\over{2}},1,{{1}\over{2}}$ and $0{\,}$, especially in 
locally-supersymmetric models [19], such as models of gauged 
supergravity [20-22].  As we have found [14] for spin-2 (graviton) 
perturbed data on the final surface $\Sigma_F{\,}$, the natural
boundary conditions are again contrasting, for odd-parity 
{\it vis}-${\it \grave a}$-{\it vis} even-parity modes.  
The remaining bosonic spin, namely $s=0$ (scalar), demands the 
existence of one or more ${\it complex}$ scalar fields (a multiplet) 
in the locally-supersymmetric models [19].  The treatment of the 
$s=0$ case in [4-7] can be replicated in the case of a complex 
scalar field $\phi{\,}$, except that the natural boundary conditions, 
consistent with the local supersymmetry, require ${\rm Re}(\phi)$ 
to be fixed at a surface such as $\Sigma_F$ (Dirichlet), whereas the 
normal derivative $\partial\bigl({\rm Im}(\phi)\bigr)/\partial n$ 
must also be fixed (Neumann).  Of course, this treatment extends to 
fermionic data $(s={{1}\over{2}}{\rm{\;} and}{\;} {{3}\over{2}})$, 
as described in [25,26].

In the gravitational-collapse model, by analogy with the simplifying
choice ${\,}\phi^{(1)}{\mid}_{\Sigma_I} = 0$ for the initial perturbative
scalar-field data, taken in [4-7], we take (for the purposes of 
exposition) the simplest Maxwell initial data at 
$\Sigma_{I}{\;}(t=0)$.  That is, we consider a negligibly weak magnetic
field outside the 'star':  the boundary conditions are
$$\eqalignno{\psi^{(o)}_{1\ell m}(0,r){\;}{\,}
&={\;}{\,}0{\quad},&(5.6)\cr
\bigl(\partial_{t}\psi^{(e)}_{1\ell m}\bigr)(0,r){\;}{\,}
&={\;}{\,}0{\quad}.&(5.7)\cr}$$
\noindent
Condition (5.6) is a Dirichlet condition on the initial odd-parity
magnetic field -- see Eqs.(5.1,4,5).  Condition (5.7) implies that 
we have an initially static even-parity multipole [38]. 

We now follow the analysis of the spin-0 field, and separate the
radial-and time-dependence.  In neighbourhoods of $\Sigma_I$ and
$\Sigma_F{\,}$, where an adiabatic approximation is valid, we can
'Fourier-expand' the variables $\psi^{(o)}_{1\ell m}(t,r)$ and
$\psi^{(e)}_{1\ell m}(t,r)$, subject to the initial conditions (5.6)
and (5.7).  By analogy with the scalar case [5], let us write
$$\eqalignno{\psi^{(o)}_{1\ell m}(t,r){\;}{\,} 
&={\;}{\,}\int^{\infty}_{-\infty}{\;}dk{\;}{\,}a^{(o)}_{1k\ell m}{\;}
\psi^{(o)}_{1k\ell}(r){\;}{\,}{{\sin(kt)}\over{\sin(kT)}}{\quad},
&(5.8)\cr
\psi^{(e)}_{1\ell m}(t,r){\;}{\,} 
&={\;}{\,}\int^{\infty}_{-\infty}{\;}dk{\;}{\,}a_{1k\ell m}^{(e)}{\;}
\psi^{(e)}_{1k\ell}(r){\;}{\,}{{\cos(kt)}\over{\sin(kT)}}{\quad},
&(5.9)\cr}$$
\noindent
where the radial functions $\{\psi^{(o)}_{1k\ell}(r)\}$ and
$\{\psi^{(e)}_{1k\ell}(r)\}$ are independent of $m{\,}$, given the
spherical symmetry of the background space-time.  Here, the
position-independent quantities $\{a^{(o)}_{1k\ell m}\}$ and
$\{a^{(e)}_{1k\ell m}\}$ are certain coefficients, with smooth 
dependence on the continuous variable $k{\,}$, which label the 
configuration of the electromagnetic field on the final surface 
$\Sigma_F{\,}$.

The radial functions $\{\psi^{(o)}_{1k\ell}(r)\}$ and
$\{\psi^{(e)}_{1k\ell}(r)\}$ each obey a regularity condition at the
'centre of symmetry' $r=0$ on the final surface $\Sigma_F{\,}$; this
requires that the corresponding (spatial) electric or magnetic field,
defined {\it via} Eqs.(5.1-5), should be smooth in a neighbourhood of
$r=0{\,}$.  As a consequence, the radial functions must be real:
$${\psi^{(o)}_{1k\ell}}^{*}(r){\;}{\,} 
={\;}{\,}\psi^{(o)}_{1,-k\ell}(r){\quad},{\qquad}{\quad}
{\psi^{(e)}_{1k\ell}}^{*}(r){\;}{\,} 
={\;}{\,}\psi^{(e)}_{1,-k\ell}(r){\quad}.\eqno(5.10)$$
\noindent
For small $r{\,}$, the radial functions should further be 
asymptotically proportional to a spherical Bessel function [40]:
$$\eqalignno{\psi^{(o)}_{1k\ell}(r){\;}{\,}
&\sim{\;}{\,}{(\rm const.)}_{o}{\;}r{\;}j_{\ell}(kr){\quad},&(5.11)\cr
\psi^{(e)}_{1k\ell}(r){\;}{\,}
&\sim{\;}{\,}({\rm const.})_{e}{\;}r{\;}j_{\ell}(kr){\quad},&(5.12)\cr}$$
\noindent
as $r\rightarrow 0_{+}{\,}$.  Also, the reality of the radial 
electric and magnetic fields implies that
$$\psi^{(o)}_{1\ell m}(t,r){\,} 
={\,}(-1)^{m}{\;}\psi^{(o)*}_{1\ell,-m}(t,r){\,},
{\quad}{\;}\psi^{(e)}_{1\ell m}(t,r){\,} 
={\,}(-1)^{m}{\;}\psi^{(e)*}_{1\ell,-m}(t,r){\;}.\eqno(5.13)$$
\noindent
This in turn implies that
$$a^{(o)}_{1k\ell m}{\;}{\,} 
={\;}{\,}(-1)^{m}{\;}{\,}a^{(o)*}_{1,-k\ell,-m}{\;},
{\qquad}{\quad}a^{(e)}_{1k\ell m}{\;}{\,}
={\;}{\,}(-1)^{m+1}{\;}{\,}a^{(e)*}_{1,-k\ell,-m}{\;}.\eqno(5.14)$$
\noindent
Since the potential (4.7), appearing in the $(t,r)$ wave equation 
(4.6), tends sufficiently rapidly to zero as
${\,}r\rightarrow\infty{\,}$, where the space-time is almost 
Schwarzschild, one has asymptotic (large-$r$) behaviour of 
$\psi^{(o)}_{1k\ell}(r)$ and $\psi^{(e)}_{1k\ell}(r)$ which is 
analogous to that in the scalar case (see Eq.(3.3) of [5]):
$$\eqalignno{\psi^{(o)}_{1k\ell}(r){\;}{\,}
&\sim{\;}{\,}\Bigl({\,}z^{(o)}_{k\ell}{\;}\exp(ik{\,}r^{*}_{s}){\;} 
+{\;}z^{(o)*}_{k\ell}{\,}\exp\bigl(-ik{\,}r^{*}_{s}\bigr)\Bigr){\quad}.
&(5.15)\cr
\psi^{(e)}_{1k\ell}(r){\;}{\,} 
&\sim{\;}{\,}\Bigl({\,}z^{(e)}_{k\ell}{\;}\exp(ik{\,}r^{*}_{s}){\;}
+{\;}z^{(e)*}_{k\ell}{\;}\exp(-ik{\,}r^{*}_{s})\Bigr){\quad}.
&(5.16)\cr}$$
\noindent
Here $\{z^{(o)}_{k\ell}\}$ and $\{z^{(e)}_{k\ell}\}$ are complex
coefficients, depending smoothly on the continuous variable $k{\,}$.
Also, as usual, $r^{*}_{s}$ is the Regge-Wheeler 'tortoise' coordinate
[13,31] for the Schwarzschild geometry.  As in the scalar case [4,5],
the inner product (normalisation) for the radial functions follows 
in the limit $R_{\infty}\rightarrow\infty{\,}$:
$$\eqalignno{\int^{R_\infty}_{0}{\;}dr{\;}{\,}
e^{a}{\;}\psi^{(o)}_{1k\ell}(r){\;}\psi^{(o)}_{1k'\ell}(r){\;}{\,}
&={\;}{\,}2\pi{\,}{\mid}z^{(o)}_{k\ell}{\mid}^{2}{\;} 
\Bigl(\delta(k,k')+{\,}\delta(k,-k')\Bigr){\;},
&(5.17)\cr
\int^{R_\infty}_{0}{\;}dr{\;}{\,}e^{a}{\;}\psi^{(e)}_{1k\ell}(r){\;}
\psi^{(e)}_{1k'\ell}(r){\;}{\,} 
&={\;}{\,}2\pi{\,}{\mid}z^{(e)}_{k\ell}{\mid}^{2}{\;}
\Bigl(\delta(k,k')+{\,}\delta(k,-k')\Bigr){\;}.&(5.18)\cr}$$

Finally, we are in a position to compute the classical Maxwell action
$S^{EM}_{\rm class}$ as a functional of the spin-1 boundary data on
the final surface $\Sigma_F{\,}$, whence (straightforwardly) the
semi-classical wave function for complexified time-interval $T{\,}$,
leading to the Lorentzian quantum amplitude or wave function.  Our
boundary conditions (5.6,7) above on the initial hypersurface
$\Sigma_I{\,}$, at time $t=0{\,}$, were designed so as to give zero
contribution from $\Sigma_I$ to the expression (4.14) for the
classical action $S^{EM}_{\rm class}{\,}$.  The contribution to 
(4.14) from $\Sigma_F$ is found, using Eqs.(5.8,9,17,18), to be
$$\eqalign{S^{EM}_{\rm class}
&\Bigl[\{a^{(o)}_{1k\ell m},{\;}a^{(e)}_{1k\ell m}\}\Bigr]\cr
&=-{\,}{{1}\over{2}}
\sum_{\ell m}{\,}{{(\ell -1)!}\over{(\ell +1)!}}
\int^{\infty}_{0}dk{\;}k
\Biggl[{\mid}z^{(o)}_{k\ell}{\mid}^{2}
\biggl({\mid}a^{(o)}_{1k\ell m}{\mid}^{2}
+{\rm Re}\Bigl(a^{(o)}_{1k\ell m}{\,}
a^{(o)*}_{1,-k\ell m}\Bigr)\biggr)\cr
&{\qquad}{\qquad}{\qquad}{\quad}+{\,}{\mid}z^{(e)}_{k\ell}{\mid}^{2}{\,}
\biggl({\mid}a^{(e)}_{1k\ell m}{\mid}^{2}
+{\rm Re}\Bigl(a^{(e)}_{1k\ell m}{\,}
a^{(e)*}_{1,-k\ell m}\Bigr)\biggr)\Biggr]{\;}
{\cot(kT)}\Bigl\bracevert_{\Sigma_{F}}{\;}.\cr}\eqno(5.19)$$
\noindent
As promised, this does now express the classical action 
$S^{EM}_{\rm class}$ as an explicit functional of suitably-chosen 
boundary data, namely, $\{a^{(o)}_{1k\ell m}\}$ and 
$\{a^{(e)}_{1k\ell m}\}$.
\end{section}

\begin{section}{Two-component spinor description of boundary data}
As mentioned in the Introduction, a more unified view of the boundary
conditions for perturbed data, as specified on the initial and final
space-like hypersurfaces $\Sigma_I$ and $\Sigma_F{\,}$, at least for the
bosonic cases of spins $s=0,1$ and $2$, can be gained with the help
of a description in terms of 2-component spinors [9,17,18].  In this
Section, we simply consider the boundary condition for $s=1$ in which
the magnetic field is fixed on a (space-like) boundary.  In Sec.7 of
[14], we briefly describe spinorially the analogous spin-2 boundary 
condition, namely that the magnetic part of the Weyl tensor [15-18] 
is specified on the boundary. 

Consider (in Lorentzian signature) a real Maxwell field strength
tensor $F_{\mu\nu}{\,}$, which can be written locally in terms of 
a vector potential $A_{\mu}{\,}$, as [Eq.(2.3)]
$$F_{\mu\nu}{\;}{\,} 
={\;}{\,}\nabla_{\mu}A_{\nu}{\,}-{\,}\nabla_{\nu}A_{\mu}{\;}{\,}
={\;}{\,}\partial_{\mu}A_{\nu}{\,}-{\,}\partial_{\nu}A_{\mu}{\quad}.
\eqno(6.1)$$
\noindent
A space-time index $\mu$ is related to a pair of spinor indices
$AA'{\;}(A=0,1;{\;}A'=0',1')$ through the (hermitian) spinor-valued
1-forms $e^{AA'}_{~~~\mu}{\,}$, defined by [9,17,18]
$$e^{AA'}_{~~~\mu}{\;}{\,}
={\;}{\,}e^{a}_{~\mu}{\;}{\,}\sigma_{a}^{~AA'}{\quad}.\eqno(6.2)$$
\noindent
Here, $e^{a}_{~\mu}$ denotes a (pseudo-) orthonormal basis of 1-forms
$(a =0,1,2,3),$ while $\sigma_{a}^{~AA'}$ denotes the Infeld-van der
Waerden translation symbols [9,17,18].  Knowledge of $F_{\mu\nu}$ at
a point is equivalent to knowledge of 
$$F_{AA'BB'}{\;}{\,}
={\;}{\,}F_{\mu\nu}{\;}e_{AA'}^{~~~~\mu}{\;}e_{BB'}^{~~~~~\nu}{\quad},
\eqno(6.3)$$
\noindent
at that point.  Here, for a real Maxwell field in a real space-time 
of Lorentzian signature, $F_{AA'BB'}{\,}$ is hermitian.  Further, 
the antisymmetry $F_{\mu\nu}{\,}={\,}F_{[\mu\nu]}$ implies that the
decomposition 
$$F_{AA'BB'}{\;}{\,}
={\;}{\,}\epsilon_{AB}{\;}\tilde\phi_{A'B'}{\;}
+{\;}\epsilon_{A'B'}{\;}\phi_{AB}{\quad},\eqno(6.4)$$
\noindent
holds, where $\epsilon_{AB}$ and $\epsilon_{A'B'}$ are the unprimed
and primed alternating spinors [17,18], while
$$\phi_{AB}{\;}{\,}
={\;}{\,}{{1}\over{2}}{\;}F_{AA'B}^{~~~~~~A'}{\;}{\,}
={\;}{\,}\phi_{BA}\eqno(6.5)$$
\noindent
is a symmetric spinor, and $\tilde\phi_{A'B'}$ is (in the real
Lorentzian case) its symmetric hermitian-conjugate spinor.  
Eq.(6.4) gives the splitting of the Maxwell field strength into 
its self-dual and anti-self-dual parts.  Knowledge of the 
3 complex components of $\phi_{AB}$ at a point is equivalent 
to knowledge of the 6 real components of $F_{\mu\nu}$ at
that point; also, the $\phi_{AB}$ are, in principle, physically
measurable, just as the $F_{\mu\nu}$ are.

The dual field strength $^{*}F_{\mu\nu}$ is defined [17,18], as in 
Eq.(2.5), to be 
$$^{*}F_{\mu\nu}{\;}{\,}
={\;}{\,}{{1}\over{2}}{\;}\eta_{\alpha\beta\mu\nu}{\;}
F^{\alpha\beta}{\quad}.\eqno(6.6)$$  

One finds that
$$F_{\mu\nu}{\,}+{\,}i{\,}^{*}F_{\mu\nu}{\;}{\,}
={\;}{\,}2{\,}\phi_{AB}{\;}\epsilon_{A'B'}{\;}
e^{AA'}_{~~~~\mu}{\;}e^{BB'}_{~~~~\nu}{\quad}.\eqno(6.7)$$
\noindent
The vacuum Maxwell field equations (2.2,4) can be combined to give
the equivalent version [17,18]
$$\nabla^{AA'}{\;}\phi_{AB}{\;}{\,}={\;}{\,}0{\;},{\qquad}{\quad}
\nabla^{AA'}{\;}{\tilde \phi}_{A'B'}{\;}{\,}={\;}{\,}0{\quad},
\eqno(6.8)$$
\noindent
where ${\;}\nabla^{AA'}{\,}={\;}e^{AA'\mu}{\;}{\,}\nabla_{\mu}{\;}$.

Here, we are interested in the decomposition of the Maxwell field
strength with respect to a space-like bounding hypersurface and its
associated unit (future-directed) normal vector $n^{\mu}$.  Define the
normal spinor
$$n^{AA'}{\;}{\,}
={\;}{\,}n^{\mu}{\;}e^{AA'}_{~~~~\mu}{\quad}.\eqno(6.9)$$
\noindent
Then the (purely spatial) electric and magnetic field vectors 
${\,}E_{k}{\,}$ and ${\,}B_{k}{\,}$ on the boundary can be 
expressed through
$$\eqalignno{E_{k}{\,}+{\,}i{\,}B_{k}
&{\;}{\,}={\;}{\,}2{\;}\phi_{AB}{\;}{\,}n^{A}_{~~B'}{\;}{\,}
e^{BB'}_{~~~~k}{\quad},&(6.10)\cr
E_{k}{\,}-{\,}i{\,}B_{k}
&{\;}{\,}={\;}{\,}2{\;}\tilde\phi_{A'B'}{\;}{\,}
n^{~~A'}_{B}{\;}{\,}e^{BB'}_{~~~~k}{\quad}.
&(6.11)\cr}$$ 
\noindent
In 4-vector language, the corresponding co-vector fields $E_{\mu}$ and
$B_{\mu}$ are defined by
$$E_{\mu}{\;}{\,}
={\;}{\,}n^{\nu}{\;}F_{\nu\mu}{\quad},
{\qquad}{\quad}B_{\mu}{\;}{\,}
={\;}{\,}n^{\nu}{\;}{\,}^{*}F_{\nu\mu}{\quad},\eqno(6.12)$$
\noindent
obeying
$$n^{\mu}{\;}E_{\mu}{\;}{\,}
={\;}{\,}0{\;}{\,}
={\;}{\,}n^{\mu}{\;}B_{\mu}{\quad}.\eqno(6.13)$$
\noindent
Next, for ${\,}\epsilon{\,}={\,}\pm 1{\;}$, define
$$\Psi^{AB}_{\epsilon}{\;}{\,} 
={\;}{\,}2{\;}\epsilon{\;}n^{A}_{~A'}{\;}{\,}n^{B}_{~~B'}{\;}{\,}
\tilde\phi^{A'B'}{\;}
+{\;}\phi^{AB}{\quad}.\eqno(6.14)$$ 
\noindent
Here, ${\,}\Psi^{AB}_{\epsilon}$ is symmetric on $A$ and $B{\,}$; 
this spinor may be re-expressed in terms of $E_k$ and $B_k{\,}$, 
on making use of the symmetry of ${\,}n_{B}^{~~B'}{\;}e^{BA'k}{\,}$ 
on its free spinor indices $B'A'$ [9].  Here, we define
$$e^{BA'k}{\;}{\,} 
={\;}{\,}h^{k\ell}{\;}{\,}e^{BA'}_{~~~~ \ell}{\quad},\eqno(6.15)$$
\noindent
where $h^{k\ell}$ is the inverse spatial metric.  The above symmetry
property then reads
$$n_{B}^{~~B'}{\;}{\,}e^{BA'k}{\;}{\,} 
={\;}{\,}n_{B}^{~~A'}{\;}{\,}e^{BB'k}{\quad}.\eqno(6.16)$$
\noindent
From Eq.(6.17), we find the decomposition
$$\Psi^{AB}_{\epsilon}{\;}{\,} 
={\;}{\,}n^{B}_{~~B'}{\;}{\,}e^{AB'k}{\;}
\biggl((\epsilon -1){\,}E_{k}{\,}-{\,}i{\,}(\epsilon +1){\,}B_{k}\biggr)
{\quad}.\eqno(6.17)$$
\noindent
In particular, our boundary condition in Secs.4,5, where the magnetic 
field (a spatial co-vector field) is fixed on each of the initial and
final space-like hypersurfaces $\Sigma_I$ and $\Sigma_F{\,}$, 
is equivalent to fixing the spinorial quantity
$$\Psi^{AB}_{+}{\;}{\,} 
={\;}{\,}-{\,}2{\,}i{\;}n^{B}_{~B'}{\;}{\,}e^{AB'k}{\;}B_{k}\eqno(6.18)$$
\noindent
on each boundary.  Note that, even though we regard $B_k$ as having 3
real components, the left-hand side of Eq.(6.19), being symmetric 
on $(AB)$, appears to have 3 complex components.  In fact, 
$\Psi^{AB}_{+}{\,}$, as defined through Eq.(6.18), obeys a further 
hermiticity requirement, appropriate for spinors in 3 Riemannian 
dimensions (that is, on the hypersurfaces $\Sigma_I$ and $\Sigma_F$) 
[9,17,18], so re-balancing matters. 

For comparison with much of the work done on black holes and their
perturbations, one needs the Newman-Penrose formalism -- an
essentially spinorial description of the geometry [41].  
Here, at present considering only unprimed spinors, a pair  
$(o^{A},{\,}\iota^{A})$ at a point is said to be a normalised dyad 
if it gives a basis for the 2-complex-dimensional vector space of 
spinors $\omega^A$ at that point, and is normalised according to
$$o_{A}{\;}\iota^{A}{\;}{\,} 
={\;}{\,}1{\;}{\,} 
={\;}{\,}-{\,}\iota_{A}{\;}o^{A}{\quad}.\eqno(6.19)$$ 
\noindent
The unprimed field strength ${\,}\phi_{AB}{\,}={\,}\phi_{(AB)}{\,}$ 
can be projected onto the dyad, to give the 3 Newman-Penrose quantities
$$\phi_{0}{\;}{\,}
={\;}{\,}\phi_{AB}{\;}o^{A}{\;}o^{B}{\quad},
{\qquad}\phi_{1}{\;}{\,} 
={\;}{\,}{{1}\over{2}}{\;}\phi_{AB}{\;}o^{A}{\;}\iota^{B}{\quad},
{\qquad}\phi_{2}{\;}{\,} 
={\;}{\,}\phi_{AB}{\;}\iota^{A}{\;}\iota^{B}{\quad}.\eqno(6.20)$$ 
\noindent
each of which is a complex scalar field (function).  Using the
Newman-Penrose formalism to describe perturbations in the background
of a rotating Kerr black-hole geometry, Teukolsky [23] derived
separable equations for the quantities ${\,}\phi_{0}{\;}{\,}(s=1)$ 
and ${\,}r^{2}{\,}\phi_{2}{\;}{\,}(s=-1)$ 
(for further review, see [24,42].)  In our non-rotating case, 
with spherically-symmetric background, the Newman-Penrose quantity 
of most interest to us, following the work of this paper, 
is ${\,}\phi_1{\,}$.  In the language of [41], ${\,}\phi_1{\,}$ has 
spin and conformal weight zero.  Its properties are best described 
in the Kinnersley null tetrad for the Schwarzschild or Kerr geometry,
in our coordinate system [43].  Here, knowledge of a null tetrad [41] 
${\,}\ell^{\mu}{\,},{\,}n^{\mu}{\,},{\,}m^{\mu}{\,},{\,}\bar m^{\mu}$ 
of vectors at a point is equivalent to knowledge of the corresponding 
normalised spinor dyad $(o^{A},\;\iota^{A})$, through the relations
$$\eqalign{l^{\mu}&{\;}\leftrightarrow{\;}o^{A}{\;}o^{A'}{\quad}, 
{\qquad}n^{\mu}{\;}\leftrightarrow{\;}\iota^{A}{\;}\iota^{A'}{\quad},\cr
m^{\mu}&{\;}\leftrightarrow{\;}o^{A}{\;}\iota^{A'}{\quad},
{\qquad}\bar m^{\mu}{\;}\leftrightarrow{\;}\iota^{A}{\;}o^{A'}{\quad}.\cr}
\eqno(6.21)$$
\noindent
Contact between this spinorial Newman-Penrose description of (spin-1)
Maxwell perturbations, and the decomposition of the linearised Maxwell
field strength $F^{(1)}_{\mu\nu}$ given in Secs.3-5 is through the
relations
$$\phi_{1}{\;}{\,}
={\;}{\,}{{{1}\over{2r^{2}}}}{\;}\sum_{\ell m}{\;}
\Bigl(\psi^{(e)}_{1\ell m}{\;}+{\,}i{\,}\psi^{(o)}_{1\ell m}\Bigr){\;}
Y_{\ell m}(\Omega){\quad}.\eqno(6.22)$$
\noindent
Thus, ${\,}r^{2}{\,}\phi_{1}$ obeys the wave equation (4.16). 
(Note that, in the language of Geroch, Held and Penrose [44], 
${\,}\phi_1$ has conformal weight zero.)  With regard to the boundary 
conditions on $\Sigma_I$ and $\Sigma_F{\,}$, when the variables 
$\psi^{(e)}_{1\ell m}$ and $\psi^{(o)}_{1\ell m}$ are being used, 
the correct boundary data (Sec.5) involve specifying 
$\psi^{(o)}_{1\ell m}$ and 
$\bigl(\partial_{t}\psi^{(e)}_{1\ell m}\bigr)$
on $\Sigma_I$ and $\Sigma_F{\,}$.

As mentioned above for spin-2 gravitational perturbations, 
a spinorial version of the 'magnetic' boundary conditions is 
summarised in [14], in a form which makes it easier to see the 
unifying features of the system of different spins. 
\end{section}

\begin{section}{Conclusion}
In this paper, we have generalised the scalar (spin-0) calculations 
of [4,5], to include the more complicated Maxwell (spin-1) case.  
For spin-1, the linearised Maxwell field splits into a part with 
even parity and a part with odd parity; a different treatment is 
needed for each of these two cases.  In both cases, the relevant 
boundary conditions involve fixing the magnetic field on the initial 
space-like boundary $\Sigma_I$ and final boundary $\Sigma_F{\,}$.  
The main result is an explicit expression (5.19) for the classical 
(linearised) Maxwell action, as a functional of the final magnetic 
field, subject to the simplifying assumption that the magnetic field 
on the initial surface $\Sigma_I$ is zero.  From this, the Lorentzian 
quantum amplitude for photon final data can be derived, as in [5]
for spin-0 perturbative final data, by taking the limit
${\,}\theta\rightarrow 0_{+}$ of ${\,}\exp(iS_{\rm class})$, 
where $S_{\rm class}$ is the action of the classical solution of the 
boundary-value problem with prescribed initial and final data, and 
with complexified time-interval $T={\mid}T{\mid}\exp(-i\theta)$, 
where $0<\theta\leq\pi/2{\,}$.  Corresponding results for graviton 
(spin-2) final data are summarised in [14].
\end{section}

\parindent = 1 pt

\begin{section}*{References}
\everypar{\hangindent\parindent}

\noindent [1] A.N.St.J.Farley and P.D.D'Eath, 
Phys Lett. B {\bf 601}, 184 (2004).          

\noindent [2] A.N.St.J.Farley and P.D.D'Eath, 
Phys Lett. B {\bf 613}, 181 (2005).          
 
\noindent [3] A.N.St.J.Farley, 
'Quantum Amplitudes in Black-Hole Evaporation', 
Cambridge Ph.D. dissertation, approved 2002 (unpublished).

\noindent [4] A.N.St.J.Farley and P.D.D'Eath, 
'Quantum Amplitudes in Black-Hole Evaporation: I. Complex Approach', 
submitted for publication (2005).

\noindent [5] A.N.St.J.Farley and P.D.D'Eath, 
'Quantum Amplitudes in Black-Hole Evaporation: II. Spin-0 Amplitude', 
submitted for publication (2005). 
      
\noindent [6] A.N.St.J.Farley and P.D.D'Eath, 
'Bogoliubov Transformations in Black-Hole Evaporation', 
submitted for publication (2005).

\noindent [7] A.N.St.J.Farley and P.D.D'Eath, 
'Vaidya Space-Time in Black-Hole Evaporation', 
submitted for publication (2005).
 
\noindent [8] R.P.Feynman and A.R.Hibbs, 
{\it Quantum Mechanics and Path Integrals}, 
(McGraw-Hill, New York) (1965).

\noindent [9]  P.D.D'Eath, {\it Supersymmetric Quantum Cosmology}, 
(Cambridge University Press, Cambridge) (1996).

\noindent [10] P.R.Garabedian, {\it Partial Differential Equations}, 
(Wiley, New York) (1964).

\noindent [11] W.McLean, 
{\it Strongly Elliptic Systems and Boundary Integral Equations}, 
(Cambridge University Press, Cambridge) (2000); 
O.Reula, 'A configuration space for quantum gravity and solutions 
to the Euclidean Einstein equations in a slab region',
Max-Planck-Institut fur Astrophysik, {\bf MPA}, 275 (1987).
    
\noindent [12] P.D.D'Eath, 
'Loop amplitudes in supergravity by canonical quantization', in 
{\it Fundamental Problems in Classical, Quantum and String Gravity}, 
ed. N. S\'anchez (Observatoire de Paris) {\bf 166} (1999), 
hep-th/9807028.

\noindent [13] T.Regge and J.A.Wheeler, 
Phys. Rev. {\bf 108}, 1063 (1957). 

\noindent [14] A.N.St.J.Farley and P.D.D'Eath, 
'Spin-1 and Spin-2 Amplitudes in Black-Hole Evaporation', 
submitted for publication (2005). 

\noindent [15] P.D.D'Eath, Phys. Rev. D {\bf 11}, 1387 (1975).

\noindent [16] K.S.Thorne, R.H.Price and D.A.Macdonald, 
{\it Black holes. The membrane paradigm.} 
(Yale University Press, New Haven, Ct.) (1986).

\noindent [17] R.Penrose and W.Rindler, {\it Spinors and Space-Time}, 
vol. 1 (Cambridge University Press, Cambridge) (1984).

\noindent [18] R.Penrose and W.Rindler, {\it Spinors and Space-Time}, 
vol. 2 (Cambridge University Press, Cambridge) (1986).

\noindent [19] J.Wess and J.Bagger, 
{\it Supersymmetry and Supergravity},
2nd. edition, (Princeton University Press, Princeton, N.J.) (1992).

\noindent [20] P.Breitenlohner and D.Z.Freedman, 
Phys. Lett. B {\bf 115}, 197 (1982).

\noindent [21] P.Breitenlohner and D.Z.Freedman, 
Ann. Phys. (N.Y.) {\bf 144}, 249 (1982).

\noindent [22] S.W.Hawking, Phys. Lett. B {\bf 126}, 175 (1983).

\noindent [23] S.A.Teukolsky, Astrophys. J. {\bf 185}, 635 (1973).

\noindent [24] J.A.H.Futterman, F.A.Handler and R.A.Matzner,   
{\it Scattering from Black Holes} 
(Cambridge University Press, Cambridge) (1988).
 
\noindent [25] A.N.St.J.Farley and P.D.D'Eath, 
'Spin-1/2 Amplitudes in Black-Hole Evaporation', 
submitted for publication (2005).

\noindent [26] A.N.St.J.Farley and P.D.D'Eath, 
'Spin-3/2 Amplitudes in Black-Hole Evaporation', in progress.

\noindent [27] S.W.Hawking and G.F.R.Ellis, 
{\it The large scale structure of space-time}, 
(Cambridge University Press, Cambridge) (1973).

\noindent [28] T.Eguchi, P.B.Gilkey and A.J.Hanson, 
Phys. Rep. {\bf 66}, 214 (1980).

\noindent [29] R.S.Ward and R.O.Wells, 
{\it Twistor Geometry and Field Theory} 
(Cambridge University Press, Cambridge) (1990).

\noindent [30] M.Nakahara, {\it Geometry, Topology and Physics}, 
2nd. edition (Institute of Physics, Bristol) (2003).

\noindent [31] C.W.Misner, K.S.Thorne and J.A.Wheeler, 
{\it Gravitation}, (Freeman, San Francisco) (1973).

\noindent [32] J.A.Wheeler, 
'Geometrodynamics and the Issue of the Final State', 
p. 317, in {\it Relativity, Groups and Topology}, 
eds. C.DeWitt and B.DeWitt (Blackie and Son, London) (1964).

\noindent [33] J.A.Wheeler, 
'Superspace and the Nature of Quantum Geometrodynamics', 
p.242, in {\it Battelle Rencontres},
eds. C.M.DeWitt and J.A.Wheeler (Benjamin, New York) (1968).
 
\noindent [34] J.Mathews, 
J. Soc. Ind. Appl. Math. {\bf 10}, 768 (1962).

\noindent [35]  J.N.Goldberg, A.J.MacFarlane, E.T.Newman, 
F.Rohrlich and E.C.G.Sudarshan, J. Math. Phys. {\bf 8}, 2155 (1967).

\noindent [36] F.J.Zerilli, Phys. Rev. D {\bf 2}, 2141 (1970).

\noindent [37] J.D.Jackson, {\it Classical Electrodynamics}, 
(Wiley, New York) (1975).

\noindent [38] C.T.Cunningham, R.H.Price and V.Moncrief, 
Astrophys. J. {\bf 224}, 643 (1978); ibid. {\bf 230}, 870 (1979).

\noindent [39] J.A.Wheeler, {\it Geometrodynamics} 
(Academic Press, New York) (1962). 

\noindent [40] M.Abramowitz and I.A.Stegun, 
{\it Handbook of Mathematical Functions}, (Dover, New York) (1964).

\noindent [41] E.T.Newman and R.Penrose, 
J. Math. Phys. {\bf 3}, 566 (1962).

\noindent [42] S.Chandrasekhar, 
{\it The Mathematical Theory of Black Holes} 
(Oxford University Press, Oxford) (1992).

\noindent [43] W.Kinnersley, J. Math. Phys {\bf 10}, 1195 (1969).
 
\noindent [44] R.Geroch, A.Held and R.Penrose, 
J. Math. Phys. {\bf 14}, 874 (1973).

\end{section}

\end{document}